\documentclass[reprint,
superscriptaddress,frontmatterverbose
,showpacs,preprintnumbers,
amsmath,amssymb,aps,pra,floatfix]{revtex4-1}
\usepackage{graphicx}
\usepackage{graphicx,amsmath,amsfonts,amssymb
	,upgreek,txfonts,color
}
\usepackage{subfigure}
\usepackage{bm}
\usepackage[colorlinks,linkcolor=blue,citecolor=blue,urlcolor=blue,breaklinks=true]{hyperref}
\begin{document}
\title{Synchronization dynamics of two nanomechanical membranes within a Fabry-Perot cavity}
\author{F. Bemani}  \email{foroudbemani@gmail.com}
\author{Ali Motazedifard} \email{motazedifard.ali@gmail.com }
\address{Department of Physics, Faculty of Science, University of Isfahan, Hezar Jerib, 81746-73441, Isfahan, Iran}
\author{R. Roknizadeh} \email{r.roknizadeh@gmail.com}
\author{M. H. Naderi} \email{mhnaderi@phys.ui.ac.ir}
\address{Department of Physics, Faculty of Science, University of Isfahan, Hezar Jerib, 81746-73441, Isfahan, Iran}
\address{Quantum Optics Group, Department of Physics, Faculty of Science, University of Isfahan, Hezar Jerib, 81746-73441, Isfahan, Iran}

\author{D. Vitali}\email{david.vitali@unicam.it}
\address{Physics Division, School of Science and Technology, University of Camerino, I-62032 Camerino (MC), Italy}
\address{INFN, Sezione di Perugia, Perugia, Italy}
\date{\today}
\begin{abstract}
Spontaneous synchronization is a significant collective behavior of weakly coupled systems. Due to their inherent nonlinear nature, optomechanical systems can exhibit self-sustained oscillations which can be exploited for synchronizing different mechanical resonators. In this paper, we explore the synchronization dynamics of two membranes coupled to a common optical field within a cavity, and pumped with a strong blue-detuned laser drive. We focus on the system quantum dynamics in the parameter regime corresponding to synchronization of the classical motion of the two membranes. With an appropriate definition of the phase difference operator for the resonators, we study synchronization in the quantum case through the covariance matrix formalism. We find that for sufficiently large driving, quantum synchronization is robust with respect to quantum fluctuations and to thermal noise up to not too large temperatures. Under synchronization, the two membranes are never entangled, while quantum discord behaves similarly to quantum synchronization, that is, it is larger when the variance of the phase difference is smaller.
\end{abstract}
\maketitle
\section{\label{sec:Sec1}Introduction}
Since the first observation of the synchronization phenomenon in two weakly coupled pendulum clocks by Huygens, various aspects of this unique phenomenon have been studied. The collective lightning of fireflies, the beating of heart cells, chemical reactions, and audience clapping are examples of this phenomenon occurring all around us \cite{Pikovsky}. Spontaneous synchronization is of great interest because it corresponds to the case in which systems synchronize their motion only due to their mutual interaction without the existence of any external driving field. Self-sustained oscillators emerging in nonlinear systems provide a suitable platform for investigating spontaneous synchronization. They possess limit cycles, which are isolated closed attractive trajectories in phase space. For a system of coupled oscillators in a limit cycle, the phase of each oscillator typically undergoes free diffusion and is in a state of maximum uncertainty, while the difference in phase between the two coupled oscillators can be locked, i.e. it has a very narrow probability distribution, and is much more robust to noise. Synchronization can also occur in chaotic systems, whenever two or more chaotic systems adjust a given property of their motion to a common behavior, due to coupling or to an external periodic or noisy force \cite{Boccaletti}. This ranges from complete agreement of
trajectories to locking of phases.

The problem of synchronization of quantum systems has been considered more recently, from different theoretical perspectives: clock synchronization by means of quantum and classical communication protocols \cite{Chuang,Jozsa,Giovannetti2,Giovannetti}, synchronization in oscillator networks \cite{new7,new8}, synchronization of a quantum tunneling system to an external driving \cite{Goychuk}, quantum behavior of classically synchronized systems \cite{Zhirov1,Zhirov2,Zhirov3}, quantum synchronization of van der Pol oscillators \cite{Lee,Lee2,Walter,Walter2,Lorch0,Weiss}, and between two atomic ensembles \cite{Holland}. The study of synchronization in quantum systems presents additional difficulties because complete synchronization is impossible due to the uncertainty principle, while phase synchronization is nontrivial due to the controversial nature of the quantum phase operator \cite{Ban,Luis}. However, Ref. \cite{Mari1} has recently afforded the problem and suggested to describe synchronization in terms of appropriate quantum variances, and here we will further elaborate along this line. Moreover Refs. \cite{Mari1,MaxLudwig} suggested optomechanical systems as promising platforms for the investigation of synchronization at the quantum level.

In optomechanical systems (OMSs) electromagnetic radiation is coupled to one ore more mechanical oscillators (MOs) \cite{Aspelmeyer}. Suspended mirrors \cite{Aspelmeyer}, photonic crystal cavities \cite{Aspelmeyer,Eichenfield1,Eichenfield2}, levitated nanoparticles \cite{Li,Gieseler}, whispering gallery microdisks \cite{Aspelmeyer,Jiang,Wiederhecker}, ultracold atomic clouds  \cite{Aspelmeyer,Brennecke,Murch} and membrane-in-the-middle Fabry-Perot cavity systems \cite{Thompson} represent well-known examples of OMS setups.  Theoretical and experimental aspects of this emerging field of study have been investigated intensively in the last few years \cite{Aspelmeyer}. Despite their difference in the range of the parameters and their configurations, OMSs share common features. They have an inherent nonlinearity associated with the radiation pressure interaction, and a high sensitivity of the system dynamics on the detuning between the laser drive and the cavity. For some application, for instance position or force sensing \cite{Murch,Purdy,Motazedifard}, the detuning is chosen to be zero, and for some others such as backaction cooling \cite{Teufel,Chan} or state transfer \cite{Palomaki} a red-detuned laser drive is used. For entanglement purposes a blue-detuned laser is exploited \cite{Palomaki2,Paternostro,Vitali}. When an OMS is driven by a blue-detuned pump laser, radiation pressure amplifies the mechanical motion via dynamical backaction, and above a certain threshold laser power the mechanical oscillator exhibits self-sustained oscillations \cite{Marquardt1}. This phenomenon is inherently due to the non-linear nature of the optomechanical interaction. Both theoretical and experimental aspects of this phenomenon have been investigated in the classical regime \cite{Kippenberg,Carmon,Ludwig,Anetsberger,Grudinin,Zaitsev,Marquardt1,Zaitsev2,Khurgin, Khurgin2}, while in the quantum realm, limit cycles have been explored only theoretically up-to-now \cite{Vahala,Rodrigues,Armour,Qian,Nation,Dykman,Lorch}. When multiple coupled optomechanical systems and arrays are considered, new collective phenomena arise due to the mutual coupling via the radiation pressure, and in particular synchronization of limit cycles \cite{MaxLudwig,Ying2014,Heinrich,Xuereb}. Relevant experimental demonstrations of synchronization between two limit-cycle mechanical oscillators coupled to a common optical mode have been recently achieved in Refs. \cite{MianZhang,Bagheri,Shah}, while synchronization in an on-fiber optomechanical cavity to an external periodic modulation has been demonstrated in Ref. \cite{Shlomi}.

Based on these motivations, here we consider the dynamics of two membranes within a Fabry-Perot cavity with a view towards synchronization. We study the quantum dynamics of the two membranes inside the cavity, in the parameter regime where the classical dynamics manifest synchronization between them \cite{Holmes,Aveleyra}, focusing therefore on a sort of quantum analog of the original Huygens experiment. We extend the quantum measure of phase synchronization introduced in \cite{Mari1} to cover the case of two weakly coupled optomechanical systems operating in the self-sustained regime having a different amplitude. By using the Heisenberg-Langevin (HL) approach and linearizing the HL equations, we separate the deterministic dynamics and fluctuation dynamics, in order to obtain the covariance matrix (CM) to study the correlations. Defining the phase difference fluctuation operator allows us to investigate the effects of quantum fluctuations and thermal noise on synchronization, and to reveal the regimes where synchronization is obtained in the quantum realm. In particular, we find that the quantum uncertainty in the relative phase can be one order of magnitude smaller than the corresponding uncertainty in the classical case. Therefore phase synchronization in this system is robust with respect to quantum noise. Subsequently we show that at finite heat bath temperature, thermal fluctuations have a significant effect on phase synchronization in the quantum case and we also investigate whether quantum synchronization is associated with quantum correlations such as entanglement or nonzero quantum discord. In agreement with the results of Ref. \cite{Mari1}, that focused on a different model, we find that entanglement is always zero in correspondence of phase-synchronized membranes, while quantum discord appears to be a possible candidate quantum signature of synchronized limit cycles.

The paper is organized as follows. In Sec. \ref{sec:Sec2}, we describe the physical model and derive the HL equations of motion for the system operators. In Sec. \ref{sec:Sec3}, we first present and discuss the classical equations of motion and show how to synchronize two membranes in the classical regime. We then introduce the notion of phase difference in the quantum regime and examine the effect of quantum and thermal noise on the generated synchronization between the membranes. In Sec. \ref{sec:Sec4} we discuss the presence of quantum correlations i.e, entanglement and Gaussian discord, in the system. Finally, in Sec. \ref{sec:Sec5}, we present our concluding remarks.

\section{\label{sec:Sec2}System Hamiltonian and Equations of Motion}
\begin{figure}
\begin{center}
\includegraphics[width=8.5cm]{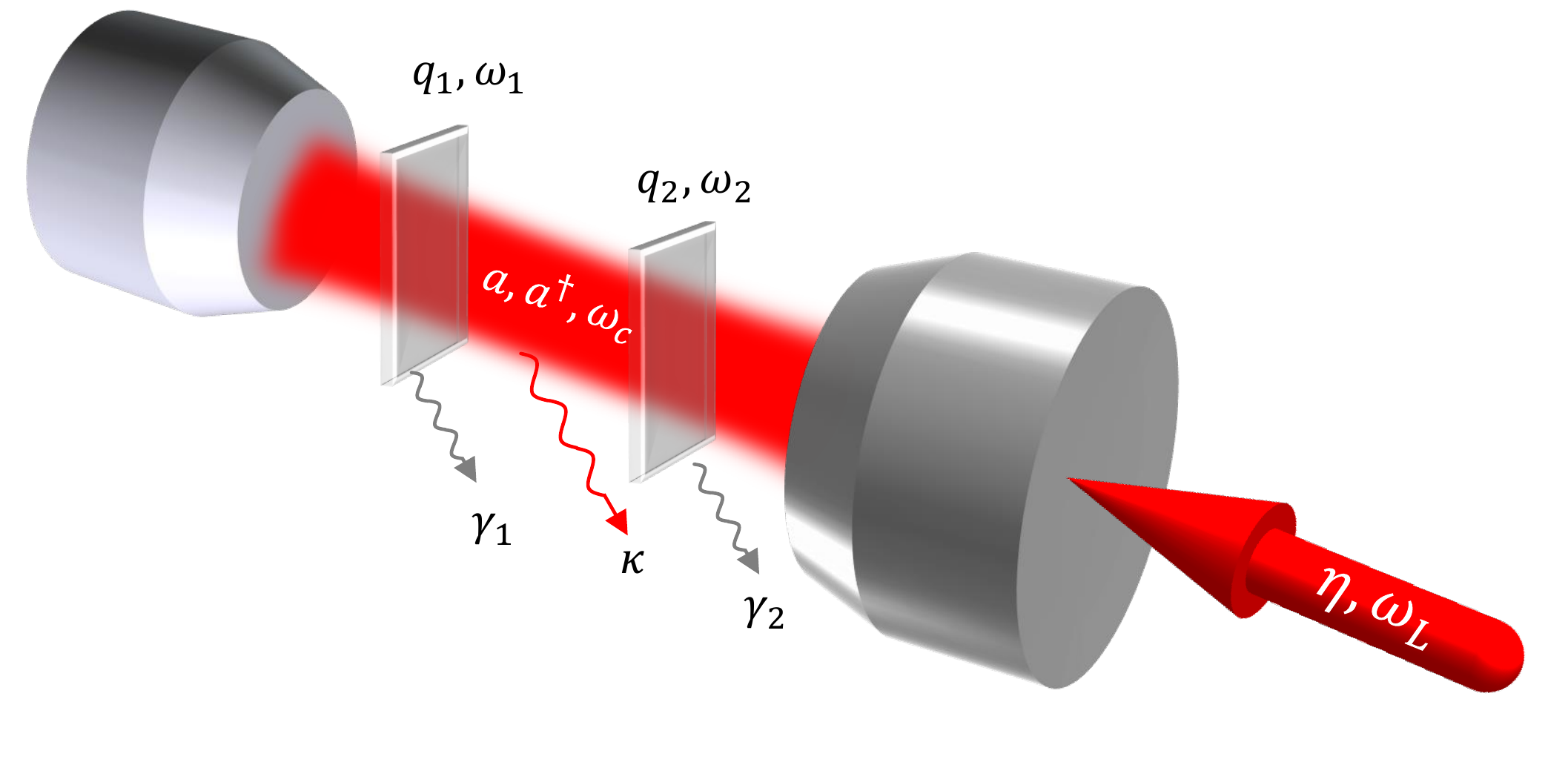}
\end{center}
\caption{Schematic illustration of a driven optical cavity containing two membranes as mechanical elements. The two membranes interacts because they are coupled to the same cavity field by the radiation pressure force. The optical cavity is pumped with a strong blue-detuned laser drive to achieve self-sustained oscillations, which can be then synchronized.}
\label{Fig:Fig1}
\end{figure}
As depicted in Fig.~\ref{Fig:Fig1}, we consider the interaction between two membranes, placed within an optical Fabry-Perot cavity. The coupling between them is through the optical field and there is no direct mechanical coupling \cite{Holmes,Aveleyra}. The Hamiltonian of the system can be written as
\begin{eqnarray}
&&H = \hbar {\omega _c}{\hat{a}^\dag }\hat{a} + \sum\limits_{j = 1}^2 {\frac{{\hbar {\omega _j}}}{2}\left( {\hat{p}_j^2 + \hat{q}_j^2} \right)}  +\sum\limits_{j = 1}^2 {\hbar {G_j}{\hat{a}^\dag }\hat{a}{\hat{q}_j}} \nonumber \\
&& \qquad +i\hbar \left( {\eta {\hat{a}^\dag }{e^{-i{\omega _L}t}} - {\eta ^*}\hat{a}{e^{  i{\omega _L}t}}} \right).
\label{Eq:Hamiltonian}
\end{eqnarray}
In this Hamiltonian the first and second terms describe the
cavity and the MOs' free Hamiltonian, respectively, the third term is the optomechanical interaction, and the last term describes the input driving by a laser with frequency $\omega _{L}$ and amplitude $\eta$.
The optical mode with frequency $\omega _c$ is described by the usual bosonic annihilation and creation operators $\hat{a}$, $\hat{a}^\dag$ satisfying the commutation relation $\left[ \hat{a},{\hat{a}^\dag} \right]=1$. The $j$th mechanical mode with frequency $\omega_j$ is described by the dimensionless position and momentum operators ${\hat{q}_j}=( \hat{b}_j+\hat{b}^\dag_j )/\sqrt{2}$ and ${\hat{p}_j}=( \hat{b}_j-\hat{b}^\dag_j )/\sqrt{2}i$ satisfying the commutation relation $ \left[ {\hat{q}_j},{\hat{p}_k} \right]=i \delta_{jk}$. The membrane-cavity coupling strength is given by $G_j=(d\omega_c/dq_j)\sqrt{\hbar/m_j\omega_j}$, where $m_j$ is the effective mass of the $j$-th MO.
\\
We then add fluctuation-dissipation processes affecting the optical and the mechanical modes, by adding for each of them the corresponding damping and noise term,  and write the following nonlinear HL equations (written in the interaction picture with respect to $\hbar {\omega _L}{a^\dag }a$)
\begin{subequations}
\label{Eq:HeisenbergLangevin}
\begin{eqnarray}
&&{\dot {\hat{a}} = \left( {i\Delta  - \kappa - i\sum\limits_{j = 1}^2 {{G_j}{\hat{q}_j}} } \right)\hat{a} + \eta  + \sqrt {2\kappa}  {\hat{a}^{in}}},\label{Eq:HeisenbergLangevin_A}\\
&&{{{\dot{ \hat{p}}}_j} =  - {\omega _j}{\hat{q}_j} - {G_j}{\hat{a}^\dag }\hat{a} - {\gamma _j}{\hat{p}_j} + \hat{\xi_j} },\label{Eq:HeisenbergLangevin_B}\\
&&{{{\dot{ \hat{q}}}_j} = {\omega _j}{\hat{p}_j}},\label{Eq:HeisenbergLangevin_C}
\end{eqnarray}
\end{subequations}
where $ \Delta =\omega _L - \omega _c$ denotes the detuning of the driving laser from the cavity resonance, $\kappa$ is the decay rate of the Fabry-Perot cavity and $\gamma_j$ is the mechanical damping rate of the $j$th membrane. The operator $\hat{a}^{in}$ denotes the vacuum optical input noise with zero mean value, satisfying the Markovian correlation functions
\begin{subequations}
\label{Eq:CorrelationFunctions}
\begin{eqnarray}
&&\left\langle {{\hat{a}^{in}}\left( t \right){\hat{a}^{in\dag }}\left( {t'} \right)} \right\rangle  = \delta \left( {t - t'} \right),
\label{Eq:CorrelationFunctions1}\\
&&\left\langle {{{\hat a}^{in\dag }}\left( t \right){{\hat a}^{in}}\left( {t'} \right)} \right\rangle  = 0,
 \label{Eq:CorrelationFunctions2}\\
&&\left\langle {{{\hat a}^{in}}\left( t \right){{\hat a}^{in}}\left( {t'} \right)} \right\rangle  = \left\langle {{{\hat a}^{in\dag }}\left( t \right){{\hat a}^{in\dag }}\left( {t'} \right)} \right\rangle  = 0,
\label{Eq:CorrelationFunctions3}
\end{eqnarray}
\end{subequations}
Each mechanical mode is coupled to its own independent thermal bath at temperature $T_j$ and it is subject to a Brownian stochastic force $\hat{\xi_j}(t)$ with zero mean value. In the limit of high mechanical quality factor, i.e., $Q_m^j = {\omega _j}/{\gamma _j} \gg 1$, the Brownian noise operator, ${\hat{\xi_j}}$, is delta-correlated \cite{Benguria,GiovannettiV}, and its symmetrized correlation function becomes
\begin{eqnarray}
&& \left\langle {\hat{\xi_j} \left( t \right)\hat{\xi_j} \left( {t'} \right) + \hat{\xi_j} \left( {t'} \right)\hat{\xi_j} \left( t \right)} \right\rangle /2  \nonumber \\
&& \qquad \qquad \quad = {\gamma _j}\left( {2{{\bar n}_j} + 1} \right)\delta \left( {t - t'} \right), \qquad (j=1,2)
\label{Eq:CorrelationFunctions4}
\end{eqnarray}
where ${\bar n_j} = {\left( {\exp \left( {\hbar {\omega _j}/{k_B}T_j} \right) - 1} \right)^{ - 1}}$ denotes the mean number of thermal phonons of the $j$th membrane at temperature $T_j$, with $k_B$ being the Boltzmann constant. Equations~(\ref{Eq:HeisenbergLangevin}) together with the correlation functions of Eqs. (\ref{Eq:CorrelationFunctions}) and (\ref{Eq:CorrelationFunctions4}) fully describe the dynamics of the system under consideration. An important feature of these sets of coupled equations is the intrinsic nonlinearity resulting from the optomechanical interaction between the cavity field and the two MOs. This nonlinearity plays a key role in achieving self-sustained oscillations for the MOs and their synchronization.

\section{\label{sec:Sec3} Dynamics of the System}
We can use the mean-field approximation in which the quantum operators are separated into $\hat O\left( t \right) =  O \left( t \right) + \delta \hat O\left( t \right)$, where $O \left( t \right)$ is the mean field describing the classical behavior of the system, and $\delta \hat O\left( t \right)$ is the quantum fluctuation with a zero mean value around the classical mean field \cite{Fong}.

\subsection{Classical dynamics}
The equations of motion for the classical mean fields form a set of nonlinear differential equations given by
\begin{subequations}
\label{Eq:ClassicalHeisenbergLangevin}
\begin{eqnarray}
&&{\dot {a}} = \left( {i\Delta  - \kappa - i\sum\limits_{j = 1}^2 {{G_j}{{q}_j}} } \right){a} + \eta,
\label{Eq:ClassicalHeisenbergLangevinA}\\
&&{{\dot{ {p}}}_j} =  - {\omega _j}{{q}_j} - {G_j}|a|^2 - {\gamma _j}{p}_j,
\label{Eq:ClassicalHeisenbergLangevinB}\\
&&{{\dot{ {q}}}_j} = {\omega _j}{{p}_j},
\label{Eq:ClassicalHeisenbergLangevinC}
\end{eqnarray}
 \label{Eq:ClassicalHeisenbergLangevin}
\end{subequations}
which are obtained by averaging Eqs.~(\ref{Eq:HeisenbergLangevin}) over classical and quantum fluctuations. This set of equations can have both static and dynamic solutions; however, here we are interested in dynamic solutions leading to self-induced oscillations, which we expect to be achieved when the cavity is driven on the blue-sideband $\Delta \simeq \omega_1$ and the driving power is large enough.
\begin{figure}
\includegraphics[width=8.5cm]{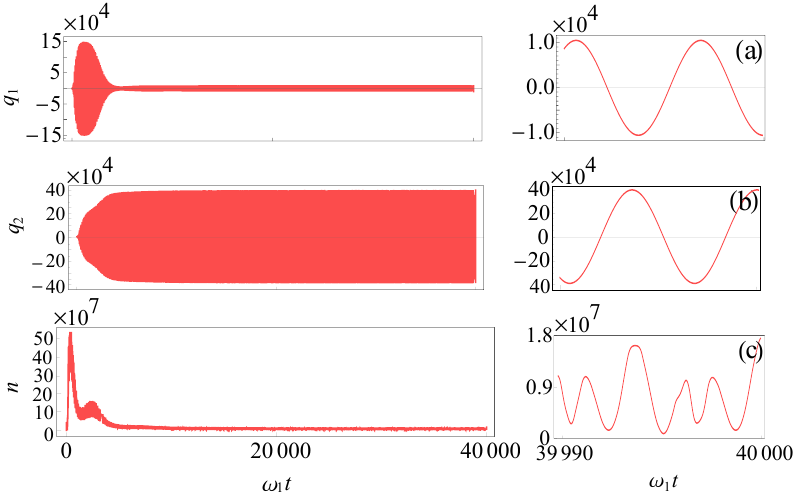}
\caption{Time evolution of the system dynamical variables vs the scaled time $\omega_1 t$ for parameters $\eta/\omega_1=3600$ and $\left(\omega_1-\omega_2 \right)/\omega_1=0.001$ (other parameters are given in the main text). (a) and (b) show the normalized position of each mechanical oscillator; (c) show the photon number inside the optical cavity. After a transient time  $\omega_1 t \sim 10^4$, the two membranes synchronize out of phase, $\phi_1-\phi_2 \simeq \pi$. }
\label{Fig:Fig2}
\end{figure}
\begin{figure}
	\includegraphics[width=8.5cm]{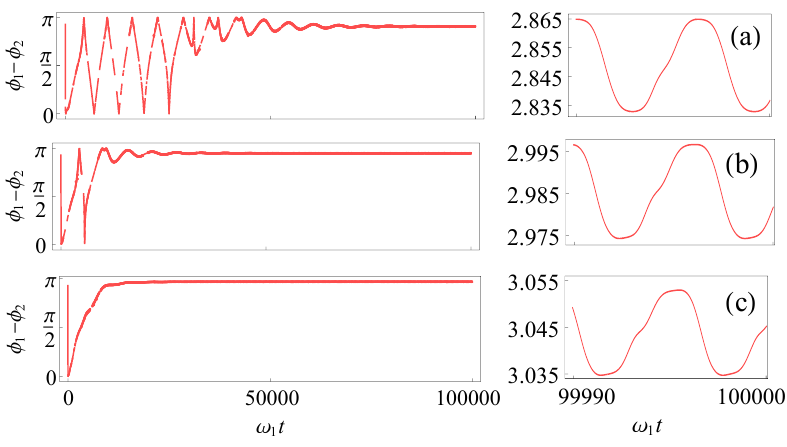}
\caption{ Time evolution of the phase difference between the two membranes vs the scaled time $\omega_1 t$ for parameters (a) $\eta/\omega_1=2000$, (b) $\eta/\omega_1=2800$, and (c) $\eta/\omega_1=3600$  (other parameters are given in the main text). For these input driving amplitudes, the two mechanical modes synchronize out of phase with a very good approximation, and $\phi_1-\phi_2 \to \pi$ for increasing $\eta/\omega_1$. }
	\label{Fig:Fig3}
\end{figure}
\begin{figure}
		\includegraphics[width=8.5cm]{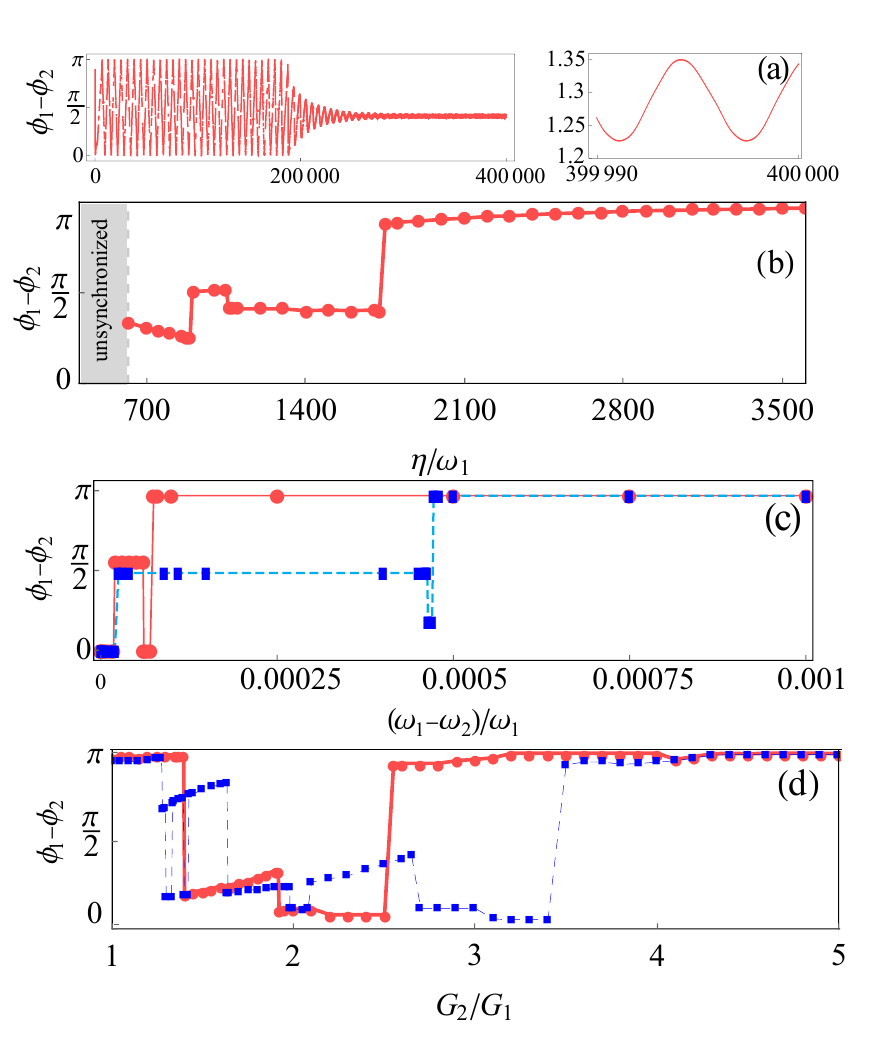}
	\caption{(a) Time evolution of the phase difference between the two membranes vs the scaled time $\omega_1 t$ for (a) $\eta/\omega_1=1200$.
(b) Stationary value of the phase difference between the two membranes with natural frequency separation $(\omega_1-\omega_2)/\omega_1=0.001$ vs $\eta/\omega_1$. There is a clear phase jump at $\eta/\omega_1 \simeq 1750$. (c) Dependence of the stationary phase difference upon the mechanical frequency separation, under two different pumping rates $\eta/\omega_1=3600$ (solid line) and $\eta/\omega_1=3000$ (dashed line). (d) Dependence of the stationary phase difference upon the ratio of optomechanical couplings, $G_2/G_1$, for $ G_1/\omega_1=10^{-5} $ under two different pumping rates $\eta/\omega_1=4000$ (solid line) and $\eta/\omega_1=3000$ (dashed line) . Other parameter are given in the main text. }
	\label{Fig:Fig4}
\end{figure}

The emergence of phase synchronization can be understood in terms of an effective Kuramoto-type equation, $\Delta \dot \phi  =  - A  - B\sin \Delta \phi  + C\cos \Delta \phi$, describing the classical dynamics of the phase difference between the two cavity-coupled MOs.  The starting point of those calculation is to consider a sinusoidal solution of the form ${q_j} = {A_j}\sin \left( {{\omega _j}t + {\phi_j^0}} \right)$ for both MOs in the self-sustained regime, and then derive an effective equation for $\Delta \phi =\phi_1^0-\phi_2^0$.  Although this ansatz will breakdown in the limit of chaotic dynamics, it is a good approximation in a large parameter region which is also experimentally achievable. Synchronization takes place after a transient time when the equation $\Delta \dot \phi=0$ has a solution otherwise synchronization cannot occur. Therefore, in order to get a synchronized system the coefficients $A$, $B$ and $C$ have to satisfy the condition $|A| \leq \sqrt{B^2+C^2}$, which implies an involved relation between system parameters, but is satisfied at large enough driving amplitude $\eta$ and not too different mechanical frequencies.

We now turn to the direct numerical investigation of the classical dynamics of the system given by Eqs.~(\ref{Eq:ClassicalHeisenbergLangevin}). From now on we will use  parameters normalized with respect to $\omega_1$, therefore, we set $\kappa/\omega_1=0.05$, $\Delta/\omega_1=1$, $\gamma_1/\omega_1=\gamma_2/\omega_1=5 \times 10^{-6}$ and $G_j/\omega_1=1 \times 10^{-5}$, which are parameters achievable in a typical setup in the resolved sideband regime \cite{Karuza}.
The time evolution of the normalized position of each MO driven by a strong blue-detuned driving laser is depicted in Fig (\ref{Fig:Fig2}a) and (\ref{Fig:Fig2}b) in the case of two membranes with a natural frequency separation $(\omega_1-\omega_2)/\omega_1=0.001$. As it can be seen, after some transient time the mechanical oscillations reach a steady state with a constant amplitude. In fact, this corresponds to self-sustained mechanical oscillations at a stable amplitude for both MOs due to nonlinear effects. Phase space trajectories of the membranes are a closed circle in this periodic steady state. It should be noted that the two MOs oscillate with different amplitudes, due to their natural frequency separation. The ratio between the two amplitudes is extremely sensitive to the frequency difference, as discussed in Ref. \cite{Holmes}, and confirmed by the plots of Fig.~\ref{Fig:Fig2}(a) and \ref{Fig:Fig2}(b). The mean photon number inside the Fabry-Perot cavity also behaves in a similar way, as shown in the bottom panel of Fig.~\ref{Fig:Fig2}(c). The time evolution of the phase difference  under three different pumping rates is shown in Fig.~\ref{Fig:Fig3}. As it can be seen, after the same transient time of Fig.~\ref{Fig:Fig2} the two membranes synchronize out of phase, i.e., $\Delta \phi \simeq \pi$. From the numerical analysis we see also that the time needed to reach the steady state depends on both the natural frequency separation of the two membranes and the pumping rate, and the results found here are consistent with the theoretical analysis of Ref. \cite{Holmes}.
Depending on the system parameters i.e., driving, frequency difference, and coupling constants $G_j$, the system under consideration can also exhibit a synchronization jump. In fact, time evolution of the phase difference between the two membranes for a smaller value of pumping rate, $\eta/\omega_1=1200$ is shown in Fig~\ref{Fig:Fig4}(a) and we see that the stationary phase difference is no longer approximately equal to $ \pi $. We can derive a sort of phase synchronization diagram by plotting the asymptotic value of the phase difference versus the input driving amplitude, as in Fig.~\ref{Fig:Fig4}(b), versus $(\omega_1-\omega_2)/\omega_1$, as shown in Fig.~\ref{Fig:Fig4}(c), and versus the ratio between the two optomechanical couplings in Fig.~\ref{Fig:Fig4}(d). We see in Fig.~\ref{Fig:Fig4}(b) that phase synchronization of the two membranes at too small driving amplitudes cannot be reached; when $(\omega_1-\omega_2)/\omega_1=0.001$, phase synchronization emerges in the system only if $\eta/\omega_1 \geq \eta_{\rm{crit}}/\omega_1=620$.
Moreover, Fig.~\ref{Fig:Fig4}(b) shows that the stationary relative phase has a sudden jump roughly from $\pi/2$ to $\pi$ at $ \eta/\omega_1\simeq 1725 $. 

The dependence of the stationary relative phase between the two membranes upon their natural frequency separation under two different pumping rates is depicted in Fig.~\ref{Fig:Fig4}(c). One has various transitions to different values of the stationary phase difference, and the results are consistent with those derived in Ref. \cite{Holmes}. Here, we set the frequency separation of the oscillators to be rather small in order to stay within the classical synchronized regime. In fact, phase synchronization is lost when the frequency difference between the membranes is too large and, as expected, the larger the driving, the larger is the maximum frequency difference for which one has phase synchronization. In particular we have numerically checked that the stationary phase difference is no longer synchronized for $ \Delta\omega/\omega_1 \geq \Delta\omega_{\rm{crit}}/\omega_1=0.00412 $ when $ \eta/\omega_1=3600$, and for $\Delta\omega/\omega_1 \geq \Delta\omega_{\rm{crit}}/\omega_1=0.00350 $ when $ \eta/\omega_1=3000$. Finally also the ratio between the two couplings $G_2/G_1$ is a critical parameter, and Fig.~\ref{Fig:Fig4}(d) shows various transitions to different values of the stationary phase difference for increasing $G_2/G_1$. Phase synchronization is no more present also if this coupling ratio is too large, i.e., the two couplings are very different. We have verified that the critical coupling ratio beyond which synchronization disappears is $ ({G_2/G_1})_{\rm{crit}}=14.72$ when $ \eta/\omega_1=3000$, and it is equal to $ ({G_2/G_1})_{\rm{crit}}=9.94$ when $ \eta/\omega_1=4000$.

\subsection{Quantum dynamics}
Here we are interested in characterizing the \emph{quantum dynamics} of the fluctuations of the system operators in the parameter regime corresponding to synchronized membranes. Reference \cite{Holmes} also afforded a preliminary investigation of such a quantum dynamics via the master equation approach, however focusing only on the output spectra and neglecting thermal fluctuations. Here, we focus on the quantum dynamics of the main signature of quantum synchronization, i.e., the variance of the phase difference operator, and adapt the approach of Ref. \cite{Mari1} to the general case in which the two MOs oscillate at different amplitudes.

In the regime of self-sustained oscillations, the amplitude and phase fluctuate around the limit cycle values $\sqrt{n_j}$ and $\phi_j$.
We can write the classical mean field ${b_j} = \sqrt {{{ n}_j}} {e^{i{{ \phi }_j}}}$, and the quantum field operator can be written as
\begin{equation}
{{\hat b}_j} = {e^{i\left( {{{ \phi }_j} + \delta {{\hat \phi }_j}} \right)}}\sqrt {{{ n}_j} + \delta {{\hat n}_j}}  \simeq \sqrt {{n_j}} {e^{i{{ \phi }_j}}}\left( {1 + i\delta {{\hat \phi }_j} + \frac{{\delta {{\hat n}_j}}}{{2{{ n}_j}}}} \right).
\end{equation}
In this representation, we have introduced the intensity fluctuation, ${\delta {{\hat n}_j}}$, and the phase fluctuation ${\delta {{\hat \phi }_j}}$ which can be easily related to the usual decomposition of the field operator in the linearized regime ${{\hat b}_j} = b_j + \delta {{\hat b}_j}$,
\begin{equation}
\delta {{\hat b}_j} = \sqrt {{{ n}_j}}e^{i\phi_j} \left( {i\delta {{\hat \phi }_j} + \frac{{\delta {{\hat n}_j}}}{{2{{n}_j}}}} \right),
\end{equation}
from which we get the following form of the phase operator fluctuations
\begin{eqnarray}
&& \delta {{\hat \phi }_j}\equiv\frac{1}{{\sqrt {{2 n_j}} }}{\delta \hat p_{{\phi _j}}}= \frac{1}{{2i\sqrt {{n_j}} }}\left( {{e^{-i{\phi _j}}}\delta {{\hat b}_j} - {e^{  i{\phi _j}}}\delta \hat b_j^\dag } \right)  \nonumber \\
&&\qquad= \frac{1}{{\sqrt {2{n_j}} }}\left( {-\sin {\phi _j}\delta {{\hat q}_j} + \cos {\phi _j}\delta {{\hat p}_j}} \right),
\end{eqnarray}
where $ \delta \hat p_{{\phi _j}} $ is a rotated momentum operator. Therefore, the fluctuation in the phase difference of the two membranes reads
\begin{equation}
\delta {{\hat \phi }_1} - \delta {{\hat \phi }_2} = \frac{{\delta {{\hat p}_{\phi _1}}}}{{\sqrt {2{{\bar n}_1}} }} - \frac{{\delta {{\hat p}_{\phi _2}}}}{{\sqrt {2{{\bar n}_2}} }}.
\end{equation}
With this in hand, one can directly use the CM formalism to calculate the variance of the fluctuation in phase difference.
\begin{figure}
\includegraphics[width=8.5cm]{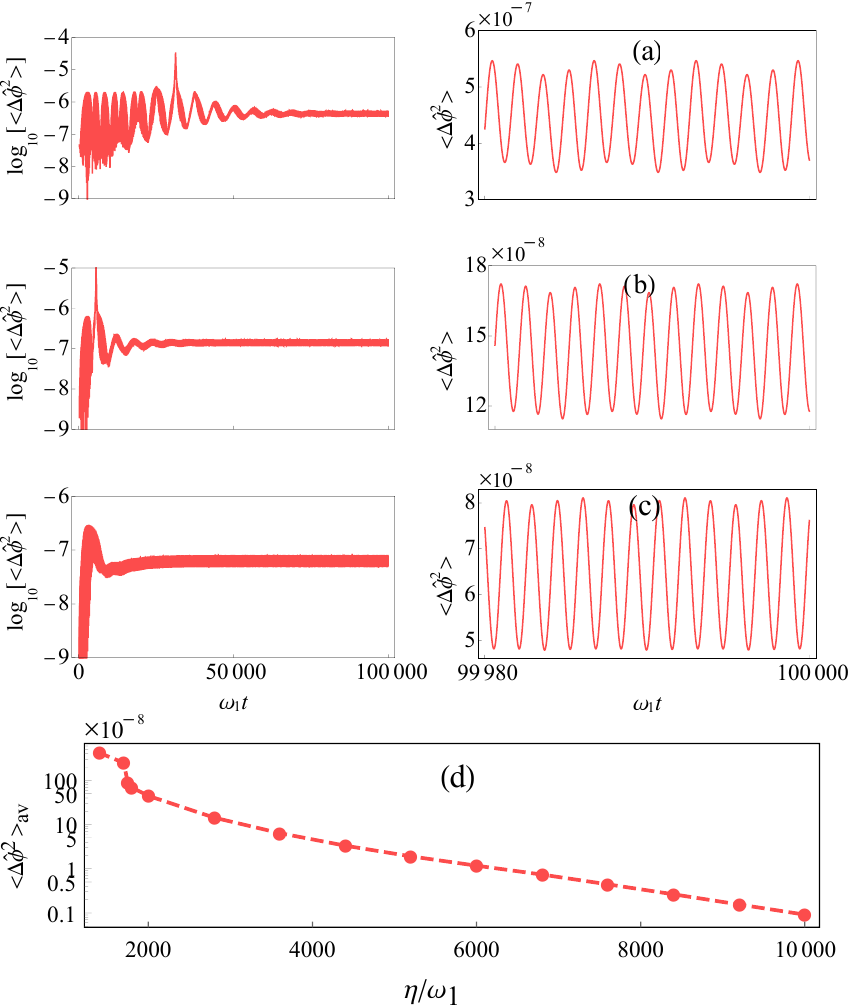}
\caption{Time evolution of the variance of the phase difference in the presence of quantum noise only ($T=0$), for different values of the optical pumping rate: (a) $\eta/\omega_1=2000$, (b) $\eta/\omega_1=2800$ and (c) $\eta/\omega_1=3600$. The stationary value of the variance remains very small, showing that synchronization is not destroyed by quantum fluctuations, and that it is more robust for larger driving rate. (d) The average of the variance of the phase difference in the self-sustained oscillation regime vs the scaled input laser power $ \eta/\omega_1$.  }
\label{Fig:Fig5}
\end{figure}

\begin{figure}
\includegraphics[width=8.5cm]{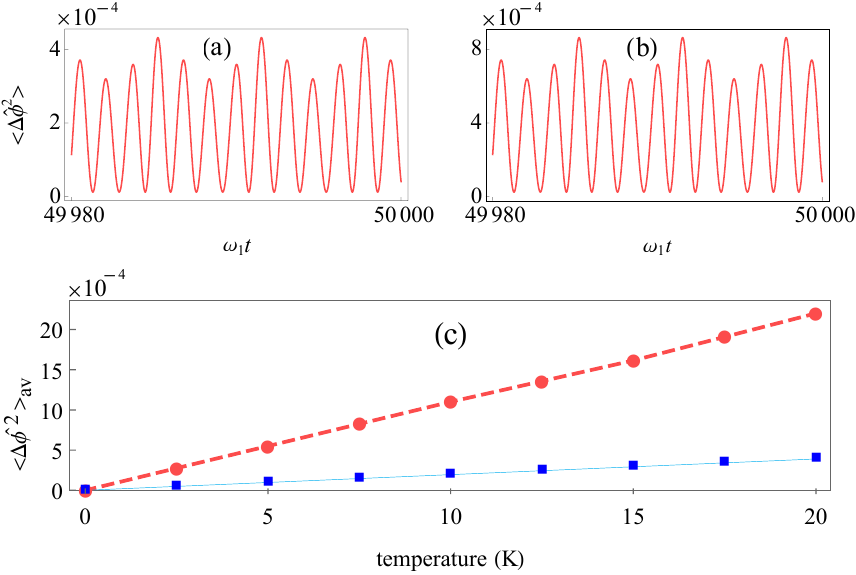}
\caption{ The effect of thermal noise on the variance of the phase difference for $\eta/\omega_1=3600$, $\omega_1=10^7~\rm{Hz}$ and (a) $T=10~\rm{K}$ (b) $T=20~\rm{K}$. (c)  The average of the variance of the phase difference in the self-sustained oscillation regime under two different pumping rates, $\eta/\omega_1=2800$ (dashed line), and $\eta/\omega_1=3600$ (solid line).}
\label{Fig:Fig6}
\end{figure}
The quantum statistical properties of the system can be investigated through the small fluctuations of the operators around the time-dependent mean values evolving according to Eqs. (\ref{Eq:ClassicalHeisenbergLangevin}). The corresponding dynamical linearized Langevin equations can be expressed in compact matrix form as
\begin{equation}
\dot u\left( t \right) = A\left( t \right)u\left( t \right) + n\left( t \right),
\label{eq:LinearizedLangevin}
\end{equation}
where we have defined the vector of fluctuation operators $u\left( t \right) = {\left( {\delta {q_1},\delta {p_1},\delta {q_2},\delta {p_2},\delta X,\delta Y} \right)^T}$ and the corresponding vector of noises $n\left( t \right) = {\left( {0,{\xi _1}\left( t \right),0,{\xi _2}\left( t \right),\sqrt \kappa  {X^{in}}\left( t \right),\sqrt \kappa  {Y^{in}}\left( t \right)} \right)^T}$. Furthermore, the drift matrix $\mathsf{A}$ is given by
\begin{equation}
\mathsf{A} = \left[ {\begin{array}{*{20}{c}}
0&{{\omega _1}}&0&0&0&0\\
{ - {\omega _1}}&{ - {\gamma _1}}&0&0&{{A_1}}&{{B_1}}\\
0&0&0&{{\omega _2}}&0&0\\
0&0&{ - {\omega _2}}&{ - {\gamma _2}}&{{A_2}}&{{B_2}}\\
{ - {B_1}}&0&{ - {B_2}}&0&{ - \kappa }&C\\
{{A_1}}&0&{{A_2}}&0&{ - C}&{ - \kappa }
\end{array}} \right],
\end{equation}
with the elements ${A_i} =  - {G_i}\sqrt 2 {\rm{Re}}\left[ a \right]$, ${B_i} =  - {G_i}\sqrt 2 {\mathop{\rm Im}\nolimits} \left[ a \right]$, and $C =  - \Delta  + \sum\limits_{j = 1}^2 {{G_j}{q_j}}$. These latter coefficients are generally time-dependent because they are the solution $a(t)$ and $q_j(t)$ of Eqs. (\ref{Eq:ClassicalHeisenbergLangevin}).
We have also used the definition of the optical mode quadratures $\delta X = \left( {\delta a + \delta {a^\dag }} \right)/\sqrt 2$ and $\delta Y = \left( {\delta a - \delta {a^\dag }} \right)/i\sqrt 2 $ together with their corresponding Hermitian noise operators ${X^{in}} \equiv \left( {{a^{in}} + {a^{in,\dag }}} \right)/\sqrt 2$ and ${Y^{in}} \equiv \left( {{a^{in}} - {a^{in,\dag }}} \right)/i\sqrt 2 $ in Eq. (\ref{eq:LinearizedLangevin}).
The evolution of the quadratures' fluctuations is described by the formal solution of Eq. (\ref{eq:LinearizedLangevin}) given by \cite{Mari1,Pinard,Mari2}
\begin{equation}
u\left( t \right) = U\left( {t,{t_0}} \right)u\left( {{t_0}} \right) + \int\limits_{{t_0}}^t {U\left( {t,s} \right)n\left( s \right)ds},
\end{equation}
in which the principal matrix solution of the homogeneous system $U\left( {t,{t_0}} \right)$ satisfies $\dot U\left( {t,{t_0}} \right) = \mathsf{A}\left( t \right)U\left( {t,{t_0}} \right)$ and $U\left( {t_0,{t_0}} \right)=1$.

In particular, the CM with entries given by ${\mathsf{V}_{ij}} \equiv \left[ {\left\langle {{u_i}\left( t  \right){u_j}\left( t  \right) + {u_j}\left( t  \right){u_i}\left( t  \right)} \right\rangle } \right]/2$ fully characterizes the mechanical and optical variances. It also includes information on the quantum correlation between the two
mechanical and the optical cavity modes. The time evolution of the CM is governed by \cite{Mari1,Mari2}
\begin{equation}
\frac{d}{{dt}}\mathsf{V}\left( t \right) = \mathsf{A}\left( t \right)\mathsf{V}\left( t \right) + \mathsf{V}\left( t \right){\mathsf{A}^T}\left( t \right) + \mathsf{D},
\end{equation}
where $\mathsf{D} = \mbox{diag} \left[0,{\gamma _1}\left( {2{{\bar n}_1} + 1} \right),0,{\gamma _2}\left( {2{{\bar n}_2} + 1} \right),\kappa ,\kappa \right]$ is the diffusion matrix. This inhomogeneous differential equation can be solved
numerically. We consider initial conditions such that both membranes are prepared in a thermal state at temperature $T$ and the cavity mode fluctuations are in the vacuum state. Therefore, the initial CM  is of the form $\mathsf{V}\left( 0 \right) = \mbox{diag} [{{\bar n}_1} + 1/2,{{\bar n}_1} + 1/2,{{\bar n}_2} + 1/2,{{\bar n}_2} + 1/2,1/2,1/2]$.

In Figs.~(\ref{Fig:Fig5}a)-(\ref{Fig:Fig5}c), we illustrate the time evolution of the variance of the phase difference in the presence of only quantum noise, i.e., in the case when $T=0$, for three different values of the optical pumping rate $\eta$. We see that when the classical dynamics corresponds to synchronized membranes, quantum noise alone is not able to destroy it: the two membranes remain essentially synchronized, with a phase difference variance which remains very small even at longer times. Moreover, the time-average of the variance of the phase difference, $\langle ( \Delta \hat{\phi})^2  \rangle_{\rm{av}}=\lim\limits_{T\rightarrow \infty}\frac{1}{T}\int_{0}^{T}\langle[\Delta \hat{\phi}(t)]^2  \rangle dt$, is shown in Fig~(\ref{Fig:Fig5}d) which states the larger the driving the smaller is the stationary value of such a phase difference variance. In order to better quantify the fact that quantum noise alone does not affect phase synchronization of the classical dynamics, we compare the quantum uncertainty in the relative phase of the membranes, i.e., $\sqrt{\langle ( \Delta \hat{\phi})^2  \rangle_{av}} $, with a classical phase uncertainty, which we take equal to the amplitude of the small residual oscillations of the phase difference at long times (see Fig.~\ref{Fig:Fig3}, right panels). In fact, the contribution of quantum noise in phase uncertainty is at least one order of magnitude smaller than the classical uncertainty.

As soon as thermal noise is included, by assuming a nonzero temperature of the membrane baths, synchronization tends to be destroyed, in the sense that the stationary value of the phase difference variance is much larger and becomes proportional to the temperature, as it typically occurs in thermal phase diffusion processes (see Fig.~\ref{Fig:Fig6}(a) and \ref{Fig:Fig6}(b) where the time evolution at two different temperatures is shown). In Fig.~\ref{Fig:Fig6}(c) we show the stationary value of the phase difference variance as a function of temperature and for two different values of the driving rate $\eta$. The linear dependence upon temperature, typical of diffusion processes, is evident, as well as the fact that the larger the optical driving, the smaller is the stationary phase difference variance. This is also expected from the fact the larger the driving, the stronger are the coherent processes induced by the radiation pressure coupling which tend to counteract the incoherent processes brought by thermal noise. Even though significantly larger than the value at zero temperature, the phase difference variance is still comparable with the classical uncertainty defined above and derived from Fig.~\ref{Fig:Fig3}, at temperature $ T\simeq4.8~\rm{K}$ and $ T\simeq20~\rm{K}$ for $ \eta/\omega_1=2800 $ and $ \eta/\omega_1=3600 $, respectively. In this sense we can say phase synchronization shows some robustness with respect to the thermal noise, at least at cryogenic temperatures.
\begin{figure}
\includegraphics[width=8.5cm]{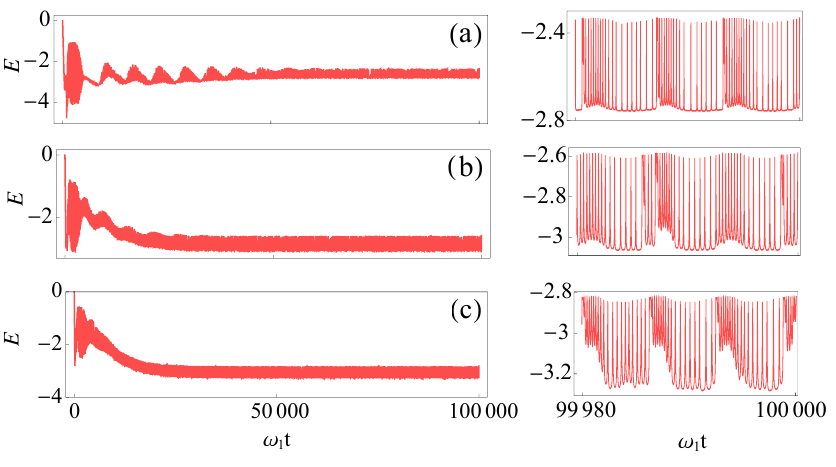}
\caption{Time evolution of $E\equiv- \ln 2{\nu_- }$, where $\nu_-$ is the smallest symplectic eigenvalue, in the self-sustained regime, at $T=0$ for (a) $\eta/\omega_1=2000$, (b) $\eta/\omega_1=2800$, and (c) $\eta/\omega_1=3600$. This quantity is always negative, showing that the two membranes are never entangled when they are phase-synchronized. }
\label{Fig:Fig7}
\end{figure}
\begin{figure}
\includegraphics[width=8.5cm]{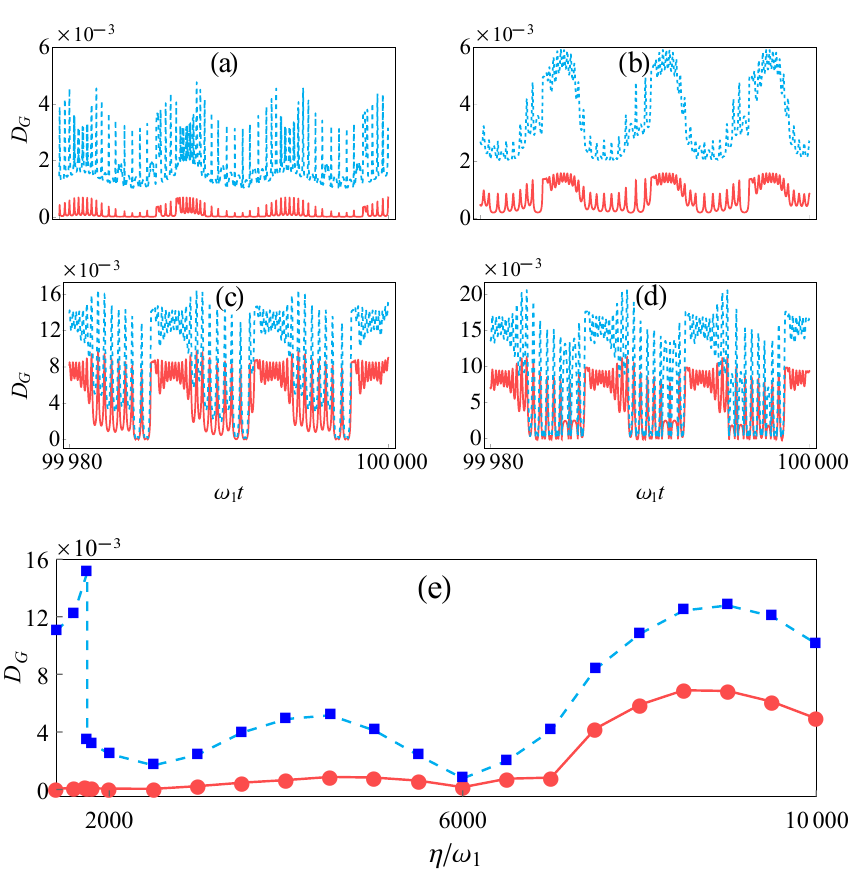}	
\caption{Time evolution of the Gaussian quantum discord in the self-sustained regime, at $T=0$. (a) $\eta/\omega_1=2800$, (b) $\eta/\omega_1=5200$, (c) $\eta/\omega_1=7600$ and (d) $\eta/\omega_1=10000$.  (e) The time-averaged Gaussian quantum discord vs the pump intensity. The red solid curves correspond to $A$-discord while the blue dashed curves correspond to the $B$-discord which can be calculated by exchanging the roles of $A$ and $B$.}	
\label{Fig:Fig8}
\end{figure}
\begin{figure}
\includegraphics[width=8.5cm]{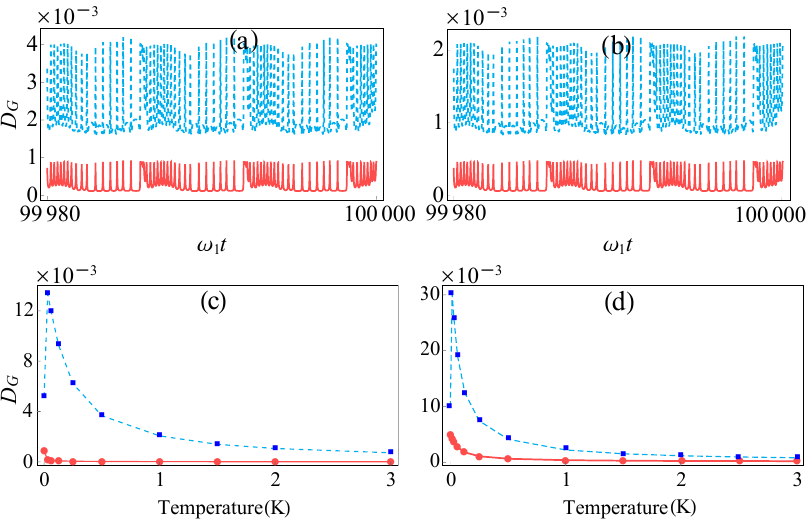}
\caption{Time evolution of the Gaussian quantum discord in the self-sustained regime under two different heat bath temperatures (a) $T=1~\rm{K}$ and (b) $T=2~\rm{K}$ , for $\eta/\omega_1=10000$. The time-averaged Gaussian quantum discord vs the heat bath temperature for two different pump intensity (c) $\eta/\omega_1=4500$ and (d) $\eta/\omega_1=10000$. Again, the solid curves correspond to the $A$ discord while the dashed curves correspond to the $B$ discord.}
\label{Fig:Fig9}
\end{figure}
\section{\label{sec:Sec4}Quantum correlations}
We now discuss the eventual presence of quantum correlations between the two membranes corresponding to a classical regime of synchronization. These correlations can be calculated from the reduced CM of the two mechanical oscillators
\begin{equation}
\mathsf{V} = \left[ {\begin{array}{*{20}{c}}
{{\mathsf{V}_{A}}}&{{\mathsf{V}_{C}}}\\
{{\mathsf{V}_{C}^T}}&{{\mathsf{V}_{B}}}
\end{array}} \right]\, ,
\end{equation}
where ${\mathsf{V}_{A}}$, ${\mathsf{V}_{B}}$, and ${\mathsf{V}_{C}}$ are $2 \times 2$  matrices. ${\mathsf{V}_{A}}$ and ${\mathsf{V}_{B}}$  account for the local properties of mechanical modes  $1$ and $2$, respectively, while  ${\mathsf{V}_{C}}$  describes intermode correlations. We quantify the degree of entanglement in terms of the logarithmic negativity, which is an entanglement monotone, and it is given by ${E_N} = \max \{ 0, E\equiv- \ln 2{\nu_- }\} $  with $ \tilde{\nu}_- = 2^{-1/2}  \left( \Sigma _-  - \sqrt {\Sigma _ - ^2 - 4\det \mathsf{V}}\right)^{1/2} $ being the smallest of the two symplectic eigenvalues of the partial transpose CM and  ${\Sigma _ \pm } = \det \mathsf{V}_{A} + \det \mathsf{V}_{B} \pm 2\det \mathsf{V}_{C}$. The time evolution of the quantity $E$ for three different values of the pumping rate is shown in Fig.~\ref{Fig:Fig7}: It is always negative and therefore the logarithm negativity is always zero even though synchronization is reached. This result is in agreement with that of Ref. \cite{Mari1} (even though for a different model in which the two resonators are directly coupled). It is then interesting to see if a weaker form of quantum correlation, quantum discord \cite{Zurek,Henderson}, is eventually present in correspondence with synchronization of the classical motion of the two membranes.
The Gaussian quantum discord of a two-mode Gaussian state is given by \cite{Adesso,Giorda}
\begin{equation}
{D_G} = f\left( {\sqrt {\beta} } \right) - f\left( {{\upsilon ^ - }} \right) - f\left( {{\upsilon ^ + }} \right) - f\left( {\sqrt \varepsilon  } \right)
\end{equation}
where
\begin{equation}
f\left( x \right) \!=\! \left( {\frac{{x \!+\! 1\!}}{2}} \right)\log_{10} \left( {\frac{{x \!+\! 1\!}}{2}} \right) \!-\! \left( {\frac{{x \!-\! 1\!}}{2}} \right) \log_{10} \left( {\frac{{x \!-\! 1\!}}{2}} \right),
\end{equation}
\begin{equation}
{\upsilon _ \pm } =\sqrt{\dfrac{{\Sigma _+ } \pm \sqrt {{\Sigma_+^{ 2}} - 4\det \mathsf{V}}}{2}  }
\end{equation}
are the two symplectic eigenvalues of the two-mode CM and
\begin{widetext}
\begin{equation}
 \varepsilon  = \left\{ {\begin{array}{*{20}{c}}
{\frac{{2{\gamma ^2} + \left( {\beta  - 1} \right)\left( {\delta  - \alpha } \right) + 2\left| \gamma  \right|\sqrt {{\gamma ^2} + \left( {\beta  - 1} \right)\left( {\delta  - \alpha } \right)} }}{{{{\left( {\beta  - 1} \right)}^2}}},}&{{{\left( {\delta  - \alpha \beta } \right)}^2} \le {{\left( {\beta  + 1} \right)}}{\gamma ^2}\left( {\alpha  + \delta } \right)};\\
{\frac{{\alpha \beta  - {\gamma ^2} + \delta  - \sqrt {{\gamma ^2} + {{\left( {\delta  - \alpha \beta } \right)}^2} - 2{\gamma ^2}\left( {\delta  + \alpha \beta } \right)} }}{{2\beta }},}&{\textrm{otherwise}},
\end{array}} \right.
 \end{equation}
 \end{widetext}
where  $\alpha  = {{\mathop{\rm det \mathsf{V}}\nolimits} _A}$, $\beta  = {{\mathop{\rm det \mathsf{V}}\nolimits} _B}$, $\gamma  = {{\mathop{\rm det \mathsf{V}}\nolimits} _C}$ and $\delta  = \det \mathsf{V}$ are  the symplectic invariants. Generally, quantum discord is intrinsically an asymmetric quantity and by swapping the roles of the two MOs, $A$ and $B$, one can obtain the $B$-discord.
The two Gaussian discords for four different pumping rates, and in the case without thermal noise, i.e., $T = 0$, are shown in Figs. \ref{Fig:Fig8}(b)-\ref{Fig:Fig8}(d). The Gaussian discord has nonzero values at times when the system classical dynamics undergoes limit cycle synchronized oscillations. This fact shows the existence of nonclassical correlations between the two mechanical oscillators, in terms of a nonzero discord, when they are phase-synchronized, and similarly to synchronization, the quantum Gaussian discord tends to increase for increasing driving rates, even though the behavior is non-monotonic. This is visible in Fig~\ref{Fig:Fig8}(e), where the time-averaged Gaussian quantum discord, $D_G^{\rm{av}}= \lim\limits_{T\rightarrow \infty}\frac{1}{T}\int_{0}^{T} D_G(t)dt $ is plotted versus the pump intensity. We also notice that the $B$ discord, the one referred to as the MO with lower frequency and typically larger oscillation amplitude, is always larger than the $A$ discord, and that the time-averaged $B$ discord has a peak in correspondence to the classical phase-synchronization jump in Fig.~\ref{Fig:Fig4}(b) at $\eta/\omega_1 \simeq 1750$. We are not able to provide an exhaustive explanation of this jump, but we observe that this is strongly reminiscent of the correspondence between classical and quantum transitions studied in Ref. \cite{Ying2014}, which focused on synchronization in a more involved system formed by two optically coupled optomechanical cavities. In such a system, the transition from in-phase to anti-phase classical synchronization has a quantum manifestation as a second-order-like phase transition of the entanglement between the two mechanical resonators in the two coupled cavities. The model studied here is simpler and does not show entanglement, as it occurs also in the model of Ref. \cite{Mari1}, but also here the sudden jump in the value of the stationary relative phase has a quantum manifestation as an abrupt change of the $B$ discord in Fig~\ref{Fig:Fig8}(e) and also as a jump in the stationary variance of the relative phase in Fig.~\ref{Fig:Fig5}(d).

In Fig.~\ref{Fig:Fig9} we show the effect of the heat bath temperature on the Gaussian discord. The $B$ discord is again always distinctly larger than the discord refereed to the higher-frequency MO; as expected, apart from a peak at very low $ T $, they both decay for increasing temperatures, but they are both non-negligible up to cryogenic temperatures.

\section{\label{sec:Sec5} concluding remarks}

We have studied the case of a membrane-in-the-middle optomechanical setup in which two membranes, interacting with the same mode of an optical Fabry-Perot cavity, can be synchronized when the cavity mode is driven with a sufficiently large power, due to the intrinsic nonlinearity of the radiation pressure interaction which leads to self-sustained oscillations. We have here focused on the dynamics of the quantum fluctuations around the synchronized classical dynamics in order to understand: (i) if there are quantum signatures of synchronization, and (ii) the robustness of these eventual signatures and of synchronization itself (quantified by the variance of the phase difference between the two mechanical oscillators) with respect to quantum and thermal noise. We have seen that, as already pointed out in Ref. \cite{Mari1}, entanglement is not related in general to synchronization, and in fact, it is absent in correspondence with synchronization of the classical motion. A more promising quantum signature of synchronization seems to be instead quantum discord. In the linearized regime of Gaussian fluctuations considered here, quantum discord is almost always nonzero as expected, but its dependence upon the relevant parameters controlling synchronization, i.e., laser driving amplitude and temperature, is always the same of the variance of the phase difference. In fact, phase synchronization and quantum discord are both robust with respect to quantum noise, and both survive in the presence of thermal noise, even though both of them decay for increasing temperatures. In conclusion the radiation pressure interaction of a sufficiently driven cavity mode is able to synchronize two membranes both coupled with the mode, and phase synchronization is also quite robust with respect to noise. As an outlook, the present scheme can be easily generalized to synchronize multiple MOs coupled to a single cavity mode.

\begin{acknowledgments}
We would like to thank the Office of Graduate Studies of 
the University of Isfahan for its support. We also gratefully
acknowledge an anonymous referee for insightful comments 
which have improved the paper. D.V. acknowledges the 
support of the European Commission through the Horizon2020-FETPROACT-2016 Project No. 732894 ``HOT.''
\end{acknowledgments}

\end{document}